\documentclass[showpacs,aps,prb,reprint,superscriptaddress]{revtex4-1}
\usepackage{bm}
\usepackage{graphicx}
\usepackage{amsmath}
\usepackage{amsfonts}
\usepackage{amssymb}
\usepackage{epstopdf}
\usepackage{hyperref}
\usepackage{float}
\hypersetup{
    colorlinks,%
    citecolor=blue,%
    linkcolor=blue,%
    urlcolor=blue
}
\usepackage{tikz}\usetikzlibrary{petri}

\begin{document}

    \newcommand{\be}   {\begin{equation}}
    \newcommand{\ee}   {\end{equation}}
    \newcommand{\ba}   {\begin{eqnarray}}
    \newcommand{\ea}   {\end{eqnarray}}
    \newcommand{\ve}  {\varepsilon}
    
    \title{ Gap oscillations and Majorana bound states in magnetic chains on superconducting honeycomb lattices }
        
    \author{Raphael L.~R.~C. Teixeira}
    \affiliation{Instituto de F\'{\i}sica, Universidade de S\~{a}o Paulo,
        C.P.\ 66318, 05315--970 S\~{a}o Paulo, SP, Brazil}
    \author{Dushko Kuzmanovski}
    \affiliation{Department of Physics and Astronomy, Uppsala University, Box 516, SE-751 20 Uppsala, Sweden}
    \author{Annica M.~Black-Schaffer}
    \affiliation{Department of Physics and Astronomy, Uppsala University, Box 516, SE-751 20 Uppsala, Sweden}
    \author{Luis G.~G.~V. Dias da Silva}
    \affiliation{Instituto de F\'{\i}sica, Universidade de S\~{a}o Paulo,
        C.P.\ 66318, 05315--970 S\~{a}o Paulo, SP, Brazil}
    
    \date{ \today }
    
    \begin{abstract}
    
        Magnetic chains on superconducting systems have emerged as a platform for realization of Majorana bound states (MBSs) in condensed matter systems with possible applications to topological quantum computation. In this work, we study the MBSs formed in  magnetic chains on two-dimensional honeycomb materials with induced superconductivity. 
        We establish chemical potential vs Zeeman splitting phase diagrams  showing that the topological regions (where MBSs appear) are strongly dependent on the spiral angle along the magnetic chain. In some of those regions, the topological phase is robust even for large values of the local Zeeman field, thus producing  topological regions which are, in a sense, ``unbounded'' in the large-field limit.
        Moreover, we show that the energy oscillations with magnetic field strength due to MBS splitting can show very different behaviors depending on the parameters. In some regimes, we find oscillations with increasing amplitudes and decreasing periods, while in the other regimes the complete opposite behavior is found.
We also find that the topological phase can become dependent on the chain length, particularly in topological regions with a very high or no upper bound. In these systems, we see a very smooth evolution from MBSs localized at chain end points to in-gap Andreev bound states spread over the full chain.          
   
    \end{abstract} 
    \maketitle

    \section{Introduction}
    \label{sec:Intro}
    
    Majorana bound states (MBSs), which naturally appear as zero-energy edge states in topological superconductors, have been receiving a great deal of attention as possible building blocks in topological quantum computation protocols \cite{Kitaev:P.U:2001}. Several theoretical  proposals to realize MBSs have been put forward, including  semiconductor nanowires with strong spin-orbit coupling \cite{Oreg:Phys.Rev.Lett.:177002:2010,Lutchyn:Phys.Rev.Lett.:77001:2010,Alicea:Reports:2012} and chains of magnetic atoms \cite{Nadj-Perge:Phys.Rev.B:020407:2013,Pawlak:NpjQuantumInformation:2:16035:2016} deposited on superconducting materials. Shortly after the predictions, promising experimental verifications were also reported \cite{Mourik:Science:1003:2012,Nadj-Perge:Science:602--607:2014}.  
    
    Interfaces of magnetic systems and topological insulators in proximity with superconductors also offer attractive possibilities in the search for MBSs in condensed-matter systems \cite{Inglot:JournalofAppliedPhysics:123709:2011,Gonzalez:Phys.Rev.B:115327:2012,Kuzmanovski:PhysRevB.94.180505,He:Phys.Rev.B:075126:2013}. Dating from the early work of Fu and Kane \cite{Fu:Phys.Rev.Lett.:96407:2008}, it has been proposed that topological superconductivity can emerge in the junction of a quantum spin Hall insulator (QSHI) supporting conventional $s$-wave superconductivity and a ferromagnetic insulator \cite{Alicea:Reports:2012}.
    Additionally, two-dimensional (2D) materials with graphenelike honeycomb structure and strong spin-orbit coupling have long been theoretically proposed as QSHIs \cite{Kane:Phys.Rev.Lett:226801:2005}.  Recently, experimental observation of QSHI behavior was reported, such as in graphene decorated with Bi$_2$Te$_3$ nanoparticles \cite{Haruyama:arXiv:2018} and in monolayer WTe$_2$ systems \cite{Wu:Science:76:2018}. These findings indicate the presence of a topological phase and support the description given by the Kane-Mele model \cite{Kane:Phys.Rev.Lett:226801:2005} in these materials. Other honeycomb lattice materials, such as silicene and stanene, are also known to have induced superconductivity when doped \cite{Chen_2013, Spencer:Springer:2016, Kuzmanovski:PhysRevB.94.180505}. This fact, coupled with their substantial spin-orbit interaction, makes all these materials very promising platforms for MBSs. 
    
    A rather simple alternative way to realize a one-dimension (1D) topological superconductor is to form a chain of magnetic impurities (defects or adatoms) on a superconducting surface (away from the edges), such that the ends of the chain might display MBSs \cite{Nadj-Perge:Phys.Rev.B:020407:2013,Pawlak:NpjQuantumInformation:2:16035:2016}. In fact, several recent experimental works using low-temperature scanning tunnel microscopy (STM) reported the presence of localized states at the end of the chains which would be consistent with MBSs  \cite{Nadj-Perge:Science:602--607:2014,Ruby:Phys.Rev.Lett.:197204:2015,Pawlak:NpjQuantumInformation:2:16035:2016,Jeon:Science:772:2017,Kim:ScienceAdvances::2018}. The main features of this arrangement can be captured by a simple (single-particle) model, considering a 1D chain of magnetic moments defined by an on-site ``Zeeman field,'' such that the ends of the chain might display MBS \cite{Nadj-Perge:Phys.Rev.B:020407:2013,Reis:Phys.Rev.B:085124:2014}. Moreover, recent experimental evidence of long-range coherent magnetic bound states in a system of diluted magnetic atoms on the surface of hexagonal 2D superconductor dichalcogenide opens interesting possibilities for producing extended Majorana quasiparticles in these systems \cite{Menard:NaturePhys:1013:2015}.
    
    In all of these platforms for realizing MBSs, there are multiple challenges even on the theoretical side. For instance, for a given system, it would be highly desirable (i) to establish the phase diagram, showing the topological phases, and (ii) to introduce protocols to experimentally differentiate MBS from any other non-topological in-gap state. 
  In the case of semiconductor nanowires, early experiments faced the challenge of distinguishing MBSs from other nontopological zero modes \cite{Franz:NatureNanotechnology:8:149:2013} such as Kondo resonances \cite{Lee:Phys.Rev.Lett.:186802:2012} and, especially, Andreev bound states (ABSs), possibly disorder induced \cite{Liu:Phys.Rev.Lett.:267002:2012}, which can mimic some behaviors of the MBSs but without topological protection. One of the theoretical proposals for establishing ``smoking-gun'' signatures is the splitting of the MBSs due to the finite-size interaction between the two MBSs on either end of the wire. This splitting, or ``gap,'' was predicted to oscillate with increasing amplitude for increasing magnetic field, while the amplitudes decay exponentially with the wire length \cite{DasSarma:Phys.Rev.B:220506:2012}. Recently,  oscillations were experimentally observed in InAs nanowires \cite{Albrecht:Nature:206:2016}, but with \textit{decreasing} amplitude with increasing field, a behavior opposite that theoretically predicted. This prompted some alternative proposals, which pointed possibly to ABSs (and not MBSs) as the source of the gap oscillations \cite{Chiu:Phys.Rev.B:054504:2017}. Several other recent studies also underlined some of the difficulties in distinguishing MBSs and ABSs \cite{Liu:Phys.Rev.B.:075161:2017,Huang:arXiv:2018}. 
     
    In this work, we focus on the distinction of MBSs and ABSs but in a different setup. Namely, we study a chain of magnetic impurities deposited on a QSHI described by the Kane-Mele model on the honeycomb lattice \cite{Kane:Phys.Rev.Lett:226801:2005} with induced superconductivity. 
 Multiple possible realizations of this simple model already exist, as discussed above \cite{Haruyama:arXiv:2018, Wu:Science:76:2018, Chen_2013, Spencer:Springer:2016,Kuzmanovski:PhysRevB.94.180505}. Even more importantly here, it is a simple system to study, yet as we will show, it has an involved phase diagram and displays complicated relationships between MBSs and ABSs. 
    An additional motivation for using this system is the possibility of exploring the role of edge QSHI states in the formation of MBSs in magnetic chains. Since QSHI edge states and MBS occur at similar energy scales, it is important to first have a firm grip on the physics of the formation of MBSs away from the edge.
    
    We show that the topological regions in the doping vs magnetic impurity field strength phase diagram are strongly influenced by the magnetic ordering along the chain, such as ferromagnetic, antiferromagnetic, or different spiral orders. 
  Especially for spiral chains, we find phase diagrams where there is no upper bound in magnetic field on the topological phase.
  In the topological regions, the presence of MBSs is very generally accompanied by  gap oscillations, but these  display surprisingly different behaviors depending on the system parameters. In some cases, a behavior similar to that predicted in Ref.\ \onlinecite{DasSarma:Phys.Rev.B:220506:2012} is obtained, with oscillation amplitude increasing and period decreasing with magnetic field strength. However, in other cases we discover the complete opposite behavior, with decreasing oscillation amplitudes and/or increasing periods. Moreover, we find that gap oscillations can also arise in nontopological regions of the phase diagram, where only ABSs are present or when the distinction between MBSs and ABSs in is not clear-cut. As an example of the latter, we show that the effective topological phase boundary can become dependent on chain length, with a smooth crossover from wire-end-point-localized MBSs to ABSs spreading over the full chain as a function of magnetic field strength. This behavior is particularly prominent in regions of the phase diagram without an upper bound in the magnetic field of the topological phase. 
  Taken together, these results show that the behavior of MBSs and their distinction from ABSs are highly parameter dependent. Ascribing particular oscillation patterns to MBSs can be treacherous, and instead, close attention has to be paid to
  system details to produce reliable predictions.
   
    The rest of the text is organized as follows: In Sec.~\ref{sec:modelmethods}, we introduce the model Hamiltonian used to describe the system and introduce the methods used to calculate its properties. The bulk system is studied in Sec.\ \ref{sec:TopPhase}, where a robust classification procedure of the bulk topological phases in terms of Majorana numbers is presented and used. We also discuss how the bulk phase diagram depends on the spiral angle of the chain and doping. Different finite-size chains and MBSs are explored in Sec.\ \ref{sec:MBS}. There we primarily focus on gap oscillations in the low-energy states and study their behavior as a function of the chain size and other parameters. We close with our conclusions, given in Sec.\ \ref{sec:Conclusions}.

    \section{Model and Method}
    \label{sec:modelmethods}
    
\begin{figure}[t]
        \begin{center}
            \begin{minipage}{.65\columnwidth}
                \includegraphics[width=1.0\linewidth]{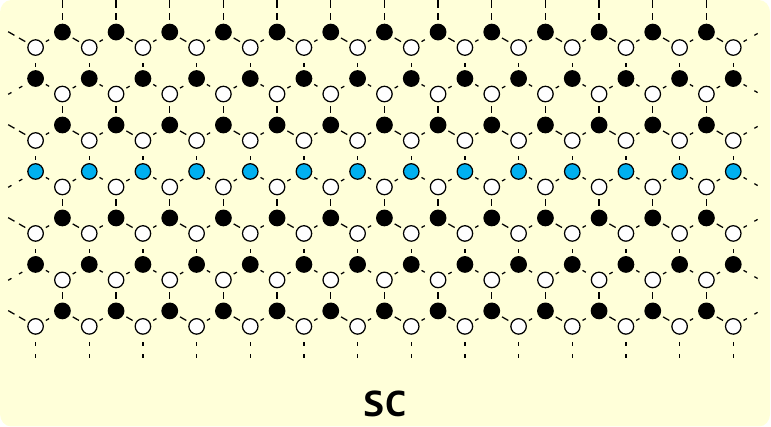}
            \end{minipage}
            \begin{minipage}{.3\columnwidth}
                \includegraphics[width=1.0\linewidth]{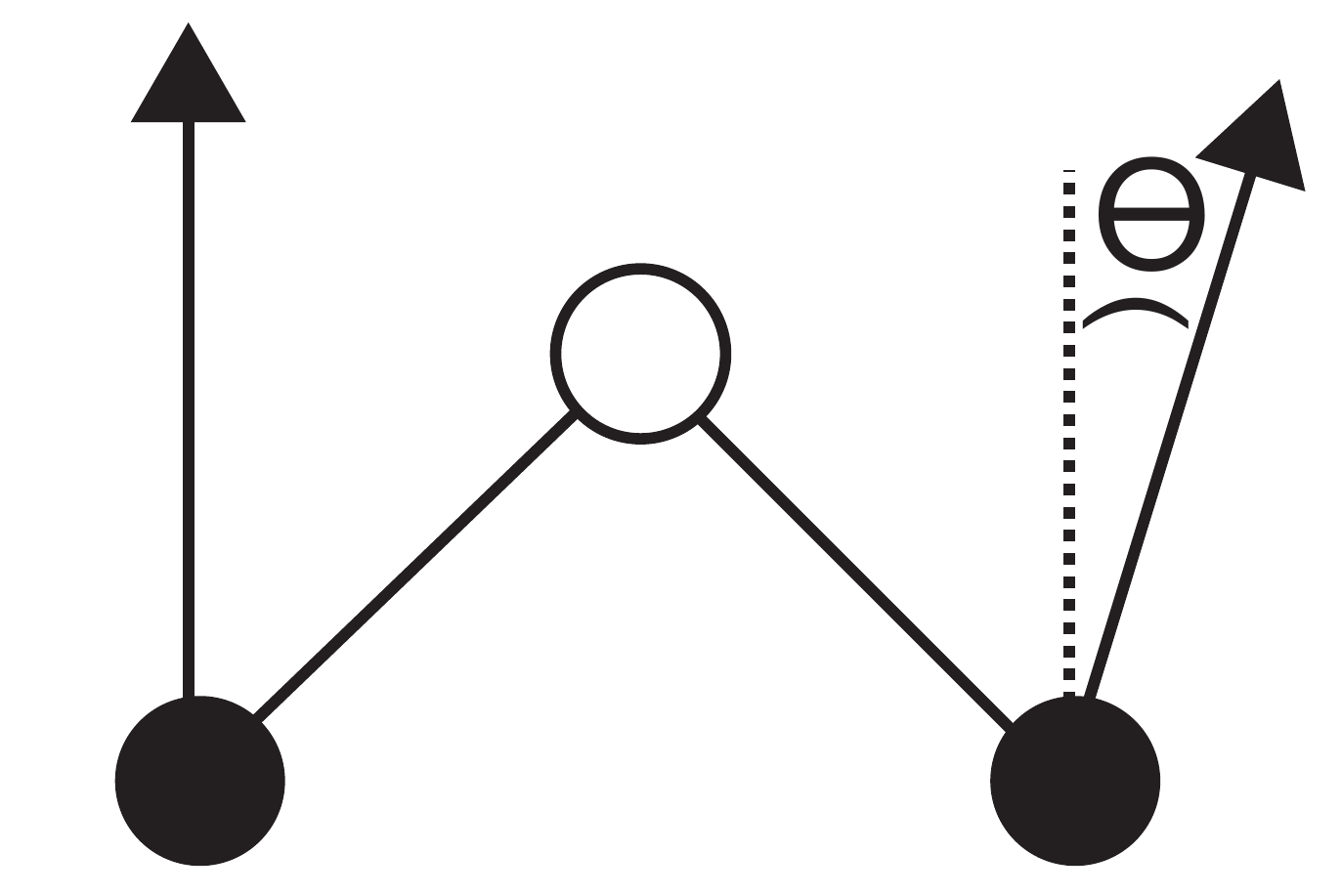}
            \end{minipage}\vspace{0.1cm}
            \begin{minipage}{1.0\columnwidth}
                \hspace{0.173\linewidth}(a)\hspace{0.433\linewidth}(b)
                \includegraphics[width=1.0\linewidth]{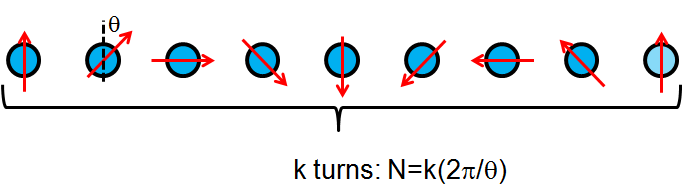}
                \hspace{0.300\linewidth}(c)
            \end{minipage}
        \end{center}
        \caption{(a) Honeycomb lattice on top of a superconducting surface (yellow) with white and black dots marking the two sublattices. Added magnetic impurities are represented as sites in blue. (b) Angle $\theta$ between the neighboring magnetic moments defines the magnetic spiral  with in-plane magnetic moments. (c) In calculations using PBCs, we impose the constraint $N\theta = 2\pi k$, where $N$ is the number of impurities and $k$ is the total number of spiral rotations. 
        }
        \label{fig.model}
    \end{figure}

    For the description of the honeycomb material on a superconducting surface, we consider the sample in Fig.~\ref{fig.model}(a), where the white and black dots represent sites belonging to the different sublattices in the honeycomb structure. The impurity chain, formed by magnetic adatoms or substitutional impurities, is represented by blue dots and is embedded in the lattice along the zigzag direction but occupies only one sublattice. This setup mimics that of impurities along a zigzag edge of the honeycomb material, although here we consider only chains fully embedded in the bulk. We fix each impurity to have a magnetic moment confined to the plane of the system ($xy$). At each site along the magnetic chain, the magnetic moment is rotated by a fixed angle $\theta$ from the preceding one, as shown in Fig.~\ref{fig.model}(b). The cases $\theta=0$ and $\theta=\pi$ thus represent ferromagnetic and antiferromagnetic ordering, respectively. A generic $\theta\neq0$ leads to a spiral magnetic order in the chain, as illustrated in Fig.~\ref{fig.model}(c).
    
To model a QSHI and also accomplish the necessary spin-orbit coupling for a topological phase even for a ferromagnetic chain, we consider the Kane-Mele Hamiltonian $\mathcal{H}_{\rm KM}$, which includes the spin-orbit coupling allowed by symmetry in  2D honeycomb systems. Combined with proximity-induced $s$-wave superconductivity and magnetic impurities, the full model Hamiltonian reads $\mathcal{H}=\mathcal{H}_{\rm KM} + \mathcal{H}_{\rm SC} + \mathcal{H}_{\rm imp}$, where
        \begin{subequations}
            \begin{multline}
            \mathcal{H}_{\rm KM} = t\sum_{\langle i,j \rangle}c^{\dagger}_{i,\sigma}c_{j,\sigma} + i\frac{\lambda_{SO}}{3 \sqrt{3}}\sum_{\langle\langle i,j \rangle\rangle} \nu_{ij} c^{\dagger}_{i,\sigma} (s_{z})_{\sigma\sigma'} c_{j,\sigma'} \\ 
            - \mu \sum_{i} c^{\dagger}_{i,\sigma} c_{i,\sigma},\label{eq.KM}
            \end{multline}\vspace{-0.5cm}
            \begin{equation}
            \mathcal{H}_{\rm SC}=-U_{\rm sc} \sum_{i}c^{\dagger}_{i,\uparrow}c_{i,\uparrow}c^{\dagger}_{i,\downarrow}c_{i,\downarrow},
            \end{equation}\vspace{-0.5cm}
            \begin{equation}
            \mathcal{H}_{\rm imp}= \sum_{i\in\mathcal{I}}V_z c^{\dagger}_{i,\sigma}(\hat{n_{i}}\cdot \vec{s})_{\sigma\sigma'}c_{i,\sigma'},
            \end{equation}
        \end{subequations}
        In the above, $t$ is the hopping between nearest neighbors in the honeycomb lattice, $\mu$ is the chemical potential, and $\lambda_{SO}$ is the spin-orbit coupling strength within each sublattice acting between next-nearest-neighbor sites. The chirality of the spin-orbit term is expressed by $\nu_{i,j}= (\bm{d}_{i} \times \bm{d}_{j})_{z}=\pm 1$, where $\bm{d}_{i,j}$ are unitary vectors connecting sites $i$ and $j$, while $s_i$ are the Pauli matrices in spin space. We assume classical spins such that the magnetic impurities are described by Zeeman-like, local magnetization terms $V_z$ with an alignment $\hat{n_{i}}(\theta) =\left( \cos\left[\theta x_{i}\right],\sin\left[\theta x_{i}\right],0\right)$, where $x_j$ is the enumeration of the chain's impurities and $\mathcal{I}$ is the set of the impurities' positions. The length of the chain is given by $\sqrt{2} \rm{a}$ times the number of impurities, with $a$ being the lattice constant.
        
        Finally, $\mathcal{H}_{\rm SC}$ represents the proximity-induced BCS-like superconductivity, given by an effective (attractive) electron-electron interaction term represented by an on-site interaction $-U_{\rm sc}$. This term effectively encodes processes of Cooper pairs leaking from the superconductor into the honeycomb lattice by adding a finite paring interaction in the honeycomb material \cite{BlackSchaffer:PhysRevB.78.024504, BlackSchaffer:PhysRevB.82.184522, BlackSchaffer:PhysRevB.83.220511}. We treat the superconductivity term in a standard mean-field approach $\mathcal{H}_{\rm SC}=\sum_{i}\Delta_{i} c^{\dagger}_{i,\uparrow} c^{\dagger}_{i,\downarrow} +{\rm H.c.}$ with $\Delta_{i}=U_{\rm sc}\left< c_{i,\uparrow} c_{i,\downarrow}\right>$, being the superconducting order parameter expressed though the self-consistency condition. This approach allow us to calculate $\Delta$ self-consistently by just assuming a constant value for $U_{\rm sc}$. This way, $\Delta$ is always appropriately adjusted, even locally, with respect to the chemical potential (or any other parameter). 

We perform the self-consistent calculations by starting with an initial guess for $\Delta$ in $\mathcal{H}$. Then we find new $\Delta$ by first diagonalizing $\mathcal{H}$ and then recalculating $\Delta$ from the self-consistency condition. We reiterate this procedure until the difference in $\Delta$ between two consecutive iterations is less than $10^{-3}$. This mean-field and self-consistency approach is clearly justified as we consider  induced superconductivity of a BCS-like superconductor at low temperature. Moreover, such proximity setup warrants only considering the two-dimensionality of the QSHI.
  
\begin{figure}[t]
        \begin{center}
            \includegraphics[width=1.0\columnwidth]{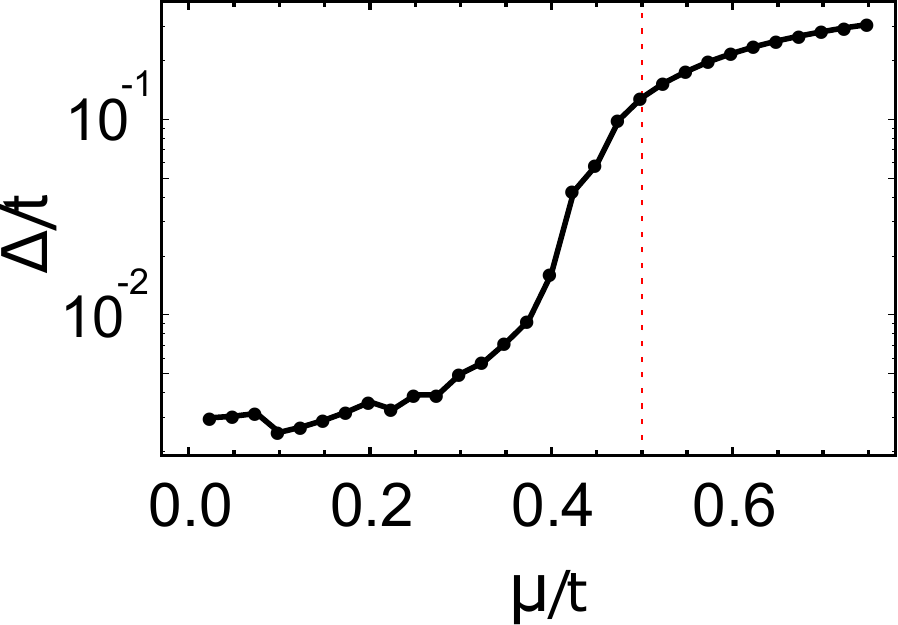}
            \caption{Superconducting order parameter $\Delta$ as a function of doping $\mu$. For small values of $\mu$,  $\Delta \sim 10^{-3}t$ but increases rapidly with doping. The vertical dashed line indicates the transition from insulator to metal. } 
            \label{fig:SOP}
        \end{center}
    \end{figure}

    All calculations presented in the paper are done for $\lambda_{\rm SO}=0.5t$, which gives a normal-state full energy gap 2$\lambda_{\rm SO}=1t$ and an electron-electron interaction $U_{\rm sc}=2t$, which is enough to yield a superconducting order parameter $\Delta\sim 10^{-3}t$ in the bulk even at small doping levels, set by a finite $\mu$. Due to our self-consistent approach, the order parameter increases with $\mu$, as shown in Fig.~\ref{fig:SOP}. As $\mu$ increases, the normal-state energy spectrum goes from that of an insulator to that of a metal, followed by an increase in $\Delta$. This, however, does not affect the existence of a topological phase and MBSs. It should be noted that, in a finite system, the order parameter is generally larger at the edges \cite{Stanescu:PhysRevB:241310:2010,Kuzmanovski:PhysRevB.96.174509}, so that these values of $\Delta$ are actually a lower bound. 
    We have checked that the resulting superconducting order parameter  $\Delta$ does not change significantly in the bulk of the sample if the self-consistency procedure is applied in a system with or without magnetic impurities. For improved numerical efficiency, we therefore calculate superconductivity self-consistently in a clean sample with no impurities, in order to obtain $\Delta$. This bulk value of $\Delta$ is then used in the system with impurities.  This procedure is further justified by the fact that features such as gap oscillations are, in general, not affected by inducing a local $\Delta$ on the magnetic impurity sites \cite{Chiu:Phys.Rev.B:054504:2017}.
    
    In order to obtain the bulk spectrum for $\mathcal{H}$, we 
 solve the resulting Bogoliubov-de Gennes equations numerically by putting the system on a torus, which is equivalent to applying periodic boundary conditions (PBCs) and sampling only at the $\Gamma$ point. In this setup, the magnetic chain spans the full length of the system. As such, the spiral angle $\theta$ and the number of chain sites $N$ are related by $N\theta = 2\pi k$, where $k$ is the total number of spiral rotations, as illustrated in Fig.~\ref{fig.model}(c). Thus, the longitudinal size of the system will depend on $\theta$, while the number of sites in the transverse (armchair) direction remains fixed, where we use 20 sites unless otherwise stated. When instead simulating finite chains fully embedded in the host, we use a supercell approach to fully isolate the chain from its periodically repeated copies, and $\theta$ is thus not constrained in this situation. Nevertheless, we used the same $\theta$ values to compare both calculations.

    \section{Topological phase diagrams}
    \label{sec:TopPhase}
    
    In order to verify the presence of a topological phase, we calculate a ``Majorana number'' for the Hamiltonian $H(N)$, where $N$ is the number of (real-space) sites of the chain using PBCs.
    In general, the Majorana number $\mathcal{M(H)}$ for a 1D Hamiltonian $\mathcal{H}$ is defined in terms of the Pfaffian (Pf) as \cite{Kitaev:P.U:2001}
    \begin{equation}
    \label{eq:M}
    \mathcal{M(H)}=\frac{\mbox{Sgn } \left[\mbox{Pf }\left(\mathcal{H}(N_1+N_2)\right)\right]}{\mbox{Sgn }\left[\mbox{Pf }\left(\mathcal{H}(N_1)\right)\right]\mbox{Sgn }\left[\mbox{Pf }\left(\mathcal{H}(N_2)\right)\right]},
    \end{equation}
    where Sgn is the sign function and $N_1$ and $N_2$ are the sizes of two chains with different lengths. It is clear that for $N_1=N_2$ we have $\mathcal{M(H)}=\mbox{Sgn} \left[\mbox{Pf }\left(\mathcal{H}(2N_1)\right)\right]$ and the Majorana number can thus be determined by a calculation  of the Pfaffian of the Hamiltonian for a chain of $2N_1$ sites.
    In our case, applying PBCs in both spatial directions of the honeycomb lattice, we arrive at a pseudo-1D system along the chain direction with a finite width set by the transverse dimension. However, as long as this finite width is kept constant, the parity of the ground state and thus the Majorana number do not change. We can therefore apply Eq.~\eqref{eq:M} to gain a topological characterization even in the case of the inhomogeneous 2D system given by $H(N)$ \cite{Reis:Phys.Rev.B:085124:2014}. 
    We notice, however, that the calculation of the Majorana number in terms of the Pfaffian is restricted to those cases when the spiral angle $\theta$ is such that the number of spiral rotations $k_1 = N_1 \theta/(2\pi)$ of the system with $N_1$ sites is an integer. In these cases, it is clear that the number of spiral rotations $k$ for $H(2N_1)$ is an even number. Thus, for a given $\theta$ and $N$, some care needs to be taken as the Pfaffian and the Majorana number can have different signs if $k$ is odd. Numerically, the Pfaffian was calculated using the package of Ref.~\onlinecite{Wimmer:ACM.Math.Softw:2012}.

    \begin{figure}[t]
        \begin{center}
              \includegraphics[width=1.0\columnwidth]{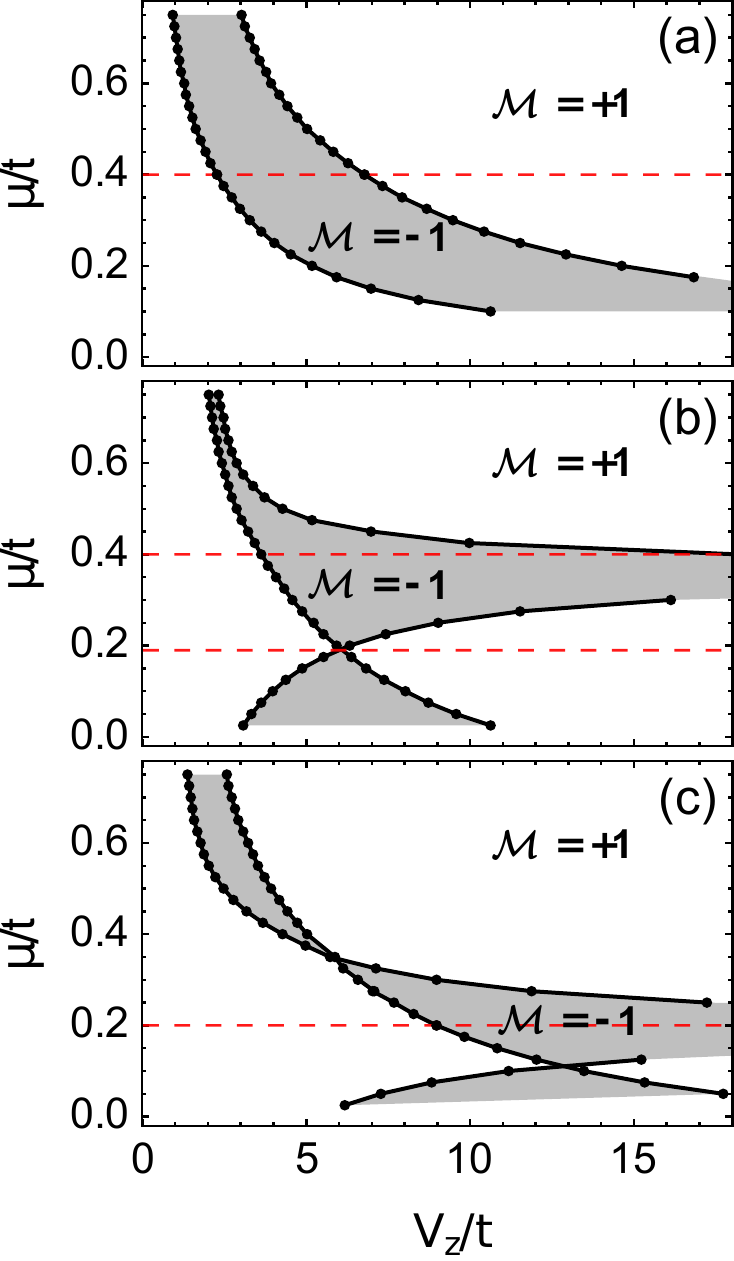}
            \caption{Phase diagram for (a) a ferromagnetic chain ($\theta=0$), (b) spiral chain with $\theta=\pi/2$, and (c) spiral chain with $\theta=\pi/4$. Topologically nontrivial phases ($\mathcal{M}=-1$) are shaded in gray, while trivial phases ($\mathcal{M}=+1$) are white. Dashed red lines correspond to the parameter space chosen in subsequent plots. }
            \label{fig:phasediagrams}
        \end{center}
    \end{figure}
    
    Having established a convenient way of deducing the existence of a topological phase, we study several different spin spiral configurations.
To this end, we consider an 800-site system with a 40-site chain for the ferromagnetic case $\theta=0$ and a 640-site system with a 32-site chain for spiral chains $\theta=\pi/4$ and $\pi/2$, always keeping the transverse direction fixed. We map the Majorana number as a function of both the chemical potential $\mu$ and Zeeman splitting $V_z$ for these three configurations, with the results summarized in Fig. \ref{fig:phasediagrams}. 
The results show that the spiral angle $\theta$ strongly influences the shape of the phase diagram. In the FM case [Fig.\ \ref{fig:phasediagrams}(a)], the boundaries between trivial and topological phases are marked by curves of the form $\mu \sim 1/V_z + \mu_{\rm min}$ such that, for a given $\mu > \mu_{\rm min}$ there is always a topological phase for some value of $V_z$. This result is similar to that found for a ferromagnetic domain at the edge of a honeycomb QSHI \cite{Kuzmanovski:PhysRevB.94.180505}.
    
    In sharp contrast we find the results for the spiral cases in Figs.~\ \ref{fig:phasediagrams}(b) and \ref{fig:phasediagrams}(c). In these cases, there are ``crossings'' of the phase boundaries such that there are discrete values of $\mu$ for which no topological phase exists. Also interestingly, there are regions where the phase boundaries become essentially horizontal. For a given $\mu$ in these regions, a lower bound exists for $V_z$ to enter a topological phase, but no upper bound exists where a trivial phase reenters. As we shall see, in these unbounded topological regions, the distinction between MBSs and ABSs is not as clear-cut as in bounded regions.  Finally, for the antiferromagnetic alignment $\theta=\pi$, we found only a trivial phase, which is consistent with previous results \cite{Reis:Phys.Rev.B:085124:2014}. We note in passing that the spiral angle is not easily controlled in experiments, which makes this distinction an even greater challenge for experimentalists.
We have also checked that for $\theta=0$ the topological phase is robust for a sublattice asymmetric spin-orbit and a next-nearest-neighbor hopping. This point is further discussed in the Appendix. For a finite spiral angle, the phase diagrams are more complicated, and some regions are somewhat enhanced, while others are suppressed, but with no qualitative changes.

    \section{Majorana bound states}
    \label{sec:MBS}
    
\begin{figure}[t]
        \begin{center}
            \includegraphics[width=1\columnwidth]{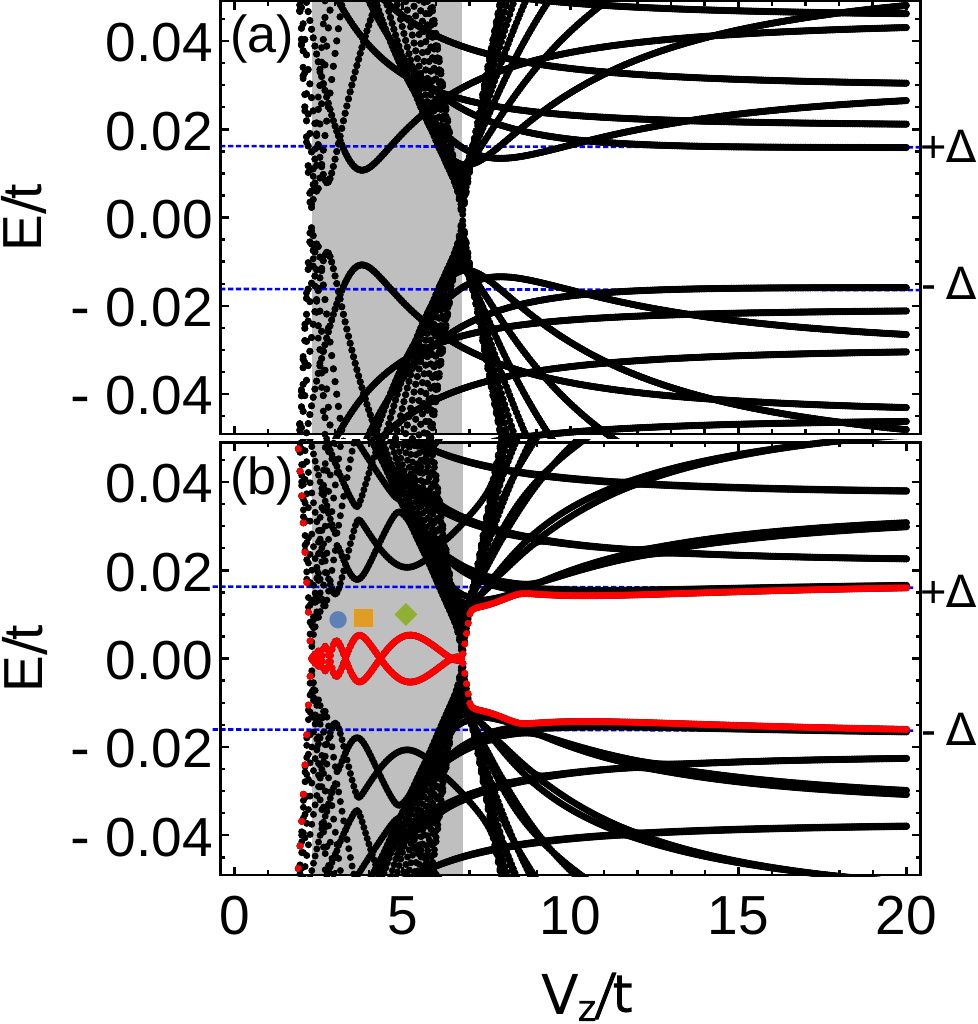}
            \caption{(a) Ferromagnetic chain ($\theta=0$) with $\mu=0.4t$ with PBCs  and (b) a fully embedded 20-site chain. States closest to zero energy are in red, horizontal dotted blue lines mark the value of the superconducting order parameter in the bulk $\pm \Delta$, and shaded regions mark the $V_z$ range of the topological phase ($\mathcal{M}=-1$) from Fig.~\ref{fig:phasediagrams}(a).
            }
            \label{fig:oscillations_mu04_theta0_vslr}
        \end{center}
    \end{figure}

    The above characterization of topological phases for bulk systems is very helpful to identify MBSs in the energy spectrum of systems which have finite magnetic chains. An example is shown in Fig.~\ref{fig:oscillations_mu04_theta0_vslr} for a ferromagnetic chain ($\theta=0$) for the parameter choices indicated by the dashed red line in Fig.~\ref{fig:phasediagrams}(a).  
    The spectrum of the system with a finite chain seen in Fig.\ \ref{fig:oscillations_mu04_theta0_vslr}(b) clearly shows low-lying edge states which are absent in the bulk calculation in Fig.\ \ref{fig:oscillations_mu04_theta0_vslr}(a). These states occur between gap closings in the bulk spectrum at  $V_z/t \approx 2$ and $V_z/t \approx 7$, values that coincide with those of the topological phase transitions for this particular value of $\mu$ in Fig.\ \ref{fig:phasediagrams}(a). These are therefore MBSs of the finite chain.
    
    We see that the MBSs oscillate as a function of the Zeeman field $V_z$, effectively opening and closing the gap. Similar gap oscillations have been identified as signatures of MBS in quantum wires where the gap has been found to change with $V_z$ according to the ansatz \cite{DasSarma:Phys.Rev.B:220506:2012}
    \begin{equation}
    \Delta \epsilon (V_z) \sim k_F(V_z) e^{-2L/\xi} \cos{\left( k_F(V_z) L \right)}.
    \label{eq:ansatz}
    \end{equation}
    Here $L$ is the chain length, $k_F(V_z) \propto V_z$ is the Fermi wavelength  and $\xi$ is the localization length of the MBS along the wire. Notably this ansatz leads to an increased amplitude and decreased period for the gap oscillations with increasing magnetic field $V_z$ and also an exponential suppression with wire length.
    \begin{figure}[H]
        \begin{center}
            \includegraphics[width=1.0\columnwidth]{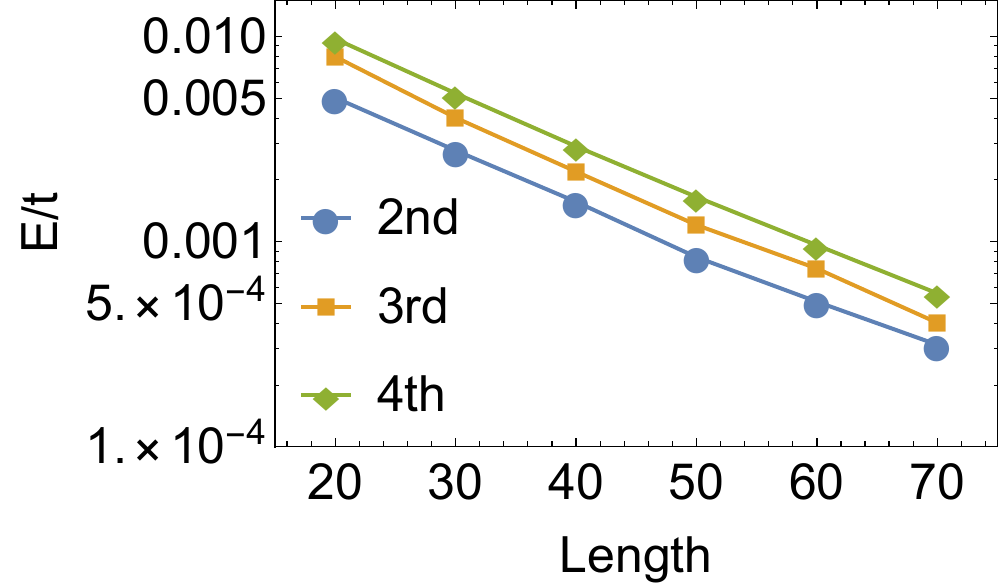}
             \caption{ Exponential decay of the amplitude of the peaks 
                marked in Fig.\ \ref{fig:oscillations_mu04_theta0_vslr} as a function of length (number of sites) of the impurity chain.}
            \label{fig:exponential_decay}
        \end{center}
    \end{figure}
    Although Eq.\ (\ref{eq:ansatz}) was originally derived for a single-band semiconductor nanowire \cite{DasSarma:Phys.Rev.B:220506:2012}, our results indicate that it can at least be qualitatively applied to describe the oscillations in $\Delta \epsilon(V_z)$ also in systems with magnetic impurities for some parameter regimes. For example, in Fig.\ \ref{fig:oscillations_mu04_theta0_vslr}(b) we show how the amplitude of the MBS energy level oscillations increases with increasing $V_z$, in agreement with the ansatz in Eq.\ (\ref{eq:ansatz}). Moreover, Fig.\ \ref{fig:exponential_decay} shows that the oscillation amplitude of the peaks marked by colored symbols in Fig.\ \ref{fig:oscillations_mu04_theta0_vslr}(b) also decreases exponentially with the chain length, in agreement with the ansatz in Eq.\ (\ref{eq:ansatz}). This is consistent with exponential topological protection for the MBS in the limit of large chains \cite{DasSarma:Phys.Rev.B:220506:2012}. It is worth mentioning that increasing the length of the chain also increases the overall number of oscillations within the topological phase ``window'' in $V_z$, which is also consistent with Eq.\ (\ref{eq:ansatz}). After the crossing at $V_z/t \approx 7$, the lowest-lying states are still subgap states inside the superconducting gap, but now localized over the whole impurity chain. However, they show no oscillatory pattern, and their energy is far from zero. Such states can therefore be identified as ABSs, which are clearly distinct from the MBSs in the topological region $2 \lesssim V_z/t \lesssim 7$ . 
    
\begin{figure}[t]
        \begin{center}
            \includegraphics[width=1\columnwidth]{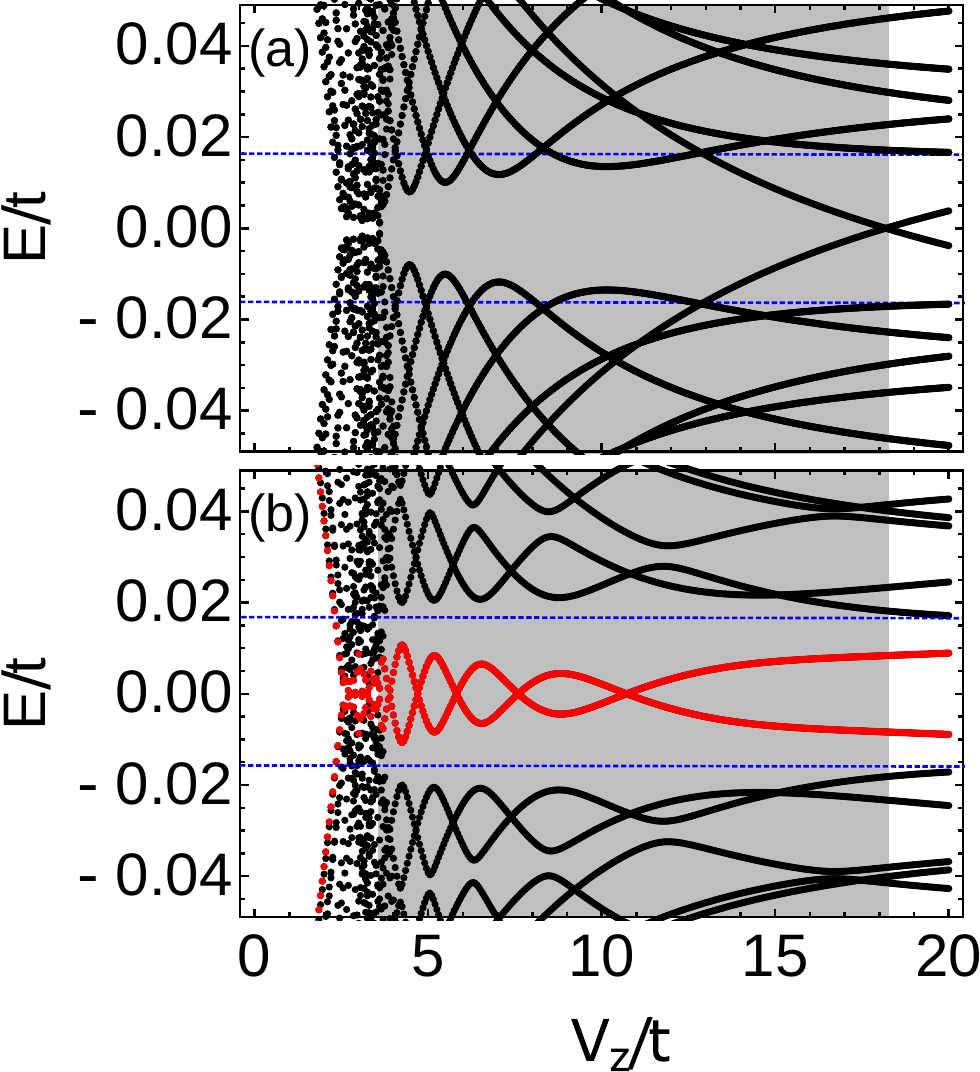}
             \caption{(a) Spiral $\theta=\pi/2$ chain with $\mu=0.4t$ with PBCs and (b) a fully embedded 21-site chain. States closest to zero energy are in red, horizontal dotted blue lines mark the value of the superconducting order parameter in the bulk $\pm \Delta$, and shaded regions mark the $V_z$ range of the topological phase ($\mathcal{M}=-1$) from Fig.~\ref{fig:phasediagrams}(b).}
            \label{fig:oscillations_mu04_thetapi2_vslr}
        \end{center}
    \end{figure}

\begin{figure}[t]
        \begin{center}
            \includegraphics[width=1\columnwidth]{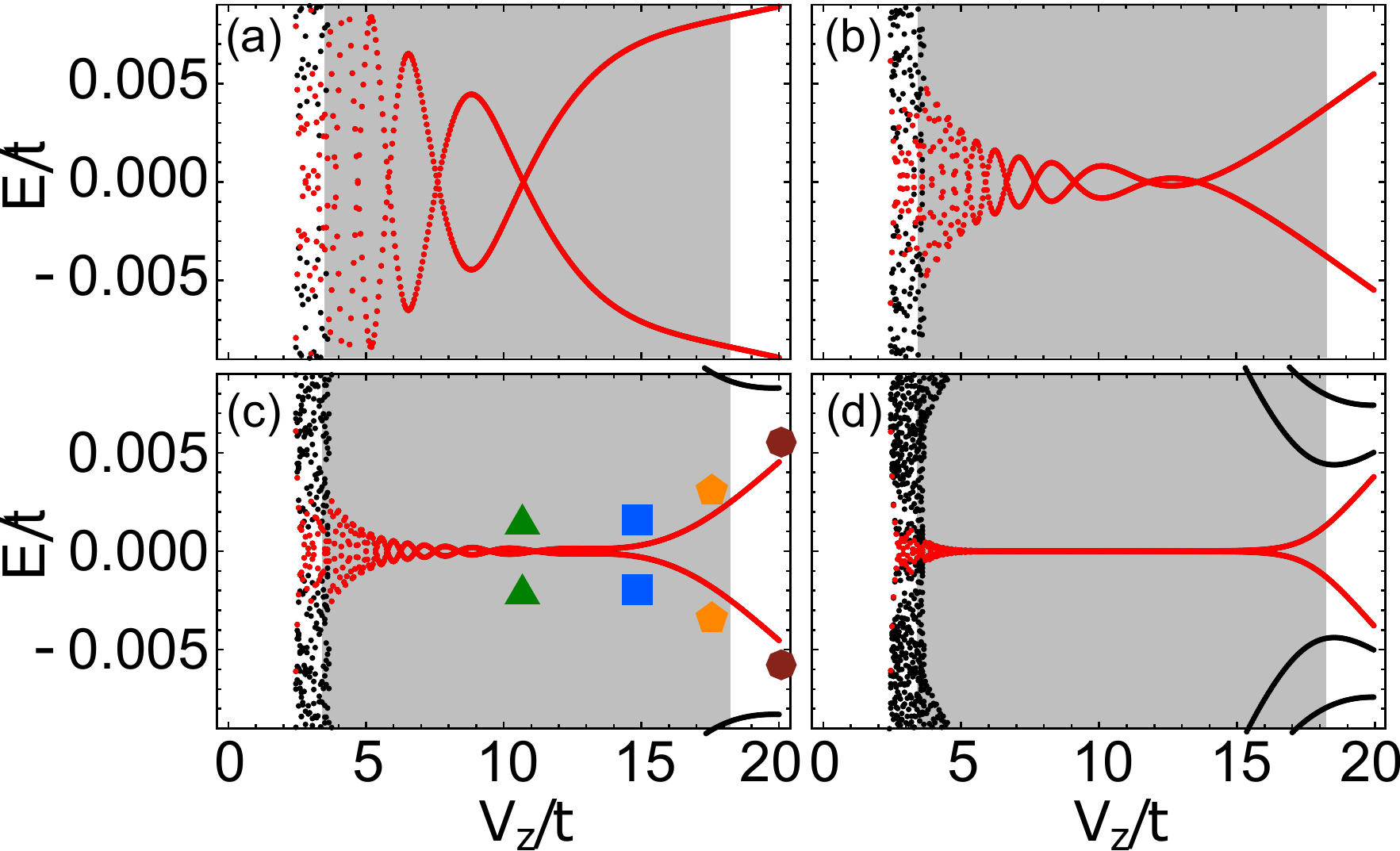}
            \caption{Low-lying spectrum of the spiral $\theta=\pi/2$ chain with $\mu=0.4t$ for (a) 21 sites, (b) 41 sites, (c) 61 sites, and (d) 121 sites. Other parameters are the same as in Fig.~\ref{fig:oscillations_mu04_thetapi2_vslr}(b).}
            \label{fig:oscillations_mu04_thetapi2}
        \end{center}
    \end{figure}

    The distinction between MBS and ABS found in Fig.~\ref{fig:oscillations_mu04_theta0_vslr} is, however, not as clear-cut in other parameter regimes. Figure \ref{fig:oscillations_mu04_thetapi2_vslr} shows results for a 21-site chain with spiral ordering ($\theta=\pi/2$) at $\mu=0.4t$. The bulk phase diagram in this case [Fig.\ \ref{fig:phasediagrams}(b)] shows that the topological region for this value of $\mu$ starts at $V_z \approx 4t$, while the upper boundary line  becomes essentially parallel to the $V_z$ axis, such that the upper bound in $V_z$ occurs at a relatively large value of $V^{\rm bulk}_{zc} \approx 18t$. These topological phase transitions are fully consistent with the bulk energy band crossings in the spectrum shown in Fig.\ \ref{fig:oscillations_mu04_thetapi2_vslr}(a). 
Comparing the bulk spectrum with that of the finite chain in Fig.\ \ref{fig:oscillations_mu04_thetapi2_vslr}(b), we find low-lying MBSs and also some important additional features. As a general trend, the energy oscillations of the MBSs deep in the topological phase actually \textit{decrease} in amplitude with increasing $V_z$. This clearly contradicts the prediction of the ansatz in Eq.\ (\ref{eq:ansatz}). We attribute this  opposite behavior to a strong hybridization between the chain and the bulk states, which is much more relevant for magnetic impurities than in nanowires. The strong hybridization is, in fact, evident when comparing  Figs.\ \ref{fig:oscillations_mu04_thetapi2_vslr}(a) and \ref{fig:oscillations_mu04_thetapi2_vslr}(b), as they show how the finite chain introduces hybridization-driven anticrossings in the spectrum that repel the bulk states at higher energies. 
The decreasing of the MBS oscillation amplitude with increasing $V_z$ remains even in the case of larger chain lengths, as shown in Fig.\ \ref{fig:oscillations_mu04_thetapi2}. While the overall amplitude of the oscillations is strongly suppressed with increasing chain size, corroborating the exponential localization of the MBSs as in the case nanowires, the oscillation amplitude always decreases with $V_z$ for a given size $L$. Thus this provides an explicit example of MBS oscillations that do not increase with magnetic field, as in often assumed.

Another important aspect shown in Fig.~\ref{fig:oscillations_mu04_thetapi2_vslr} is that the upper magnetic field for which the gap closes $V^{L}_{zc}$ is shifted downwards from $V^{\rm bulk}_{zc} \! \approx \! 18t$ to $V^{L=21}_{zc} \! \approx \! 10t$, where we find the last zero-energy crossing for the low-energy state. In fact, there is {\it no} closing of the bulk gap; only the MBSs start to very slowly approach the bulk energy gap.
This shift of the phase transition is size dependent, as shown in Fig.\ \ref{fig:oscillations_mu04_thetapi2}.  However, while the upper critical $V^{L}_{zc}$ increases with the chain length $L$, up to $V^{L=121}_{zc} \approx 15t$ for a 121-site chain, as seen in Fig.\ \ref{fig:oscillations_mu04_thetapi2}(d), it is still significantly smaller than the bulk value of $V^{\rm bulk}_{zc} \! \approx \! 18t$. 
A possible explanation for this behavior is that the low-lying MBS always shows  hybridization with the bulk states, producing a slight shift in the gap closing point and no closing of the bulk spectrum even for longer chains. In fact, we see the bulk having a tendency to a gap closing in the longest wire (black states in Fig.~\ref{fig:oscillations_mu04_thetapi2}) but not in any of the shorter wires.
The experimental implications of this result are  far reaching: The MBSs located at the chain end points exist only within the topological phase but evolve energy wise smoothly into the ABSs in the trivial phase. The transition not only is smooth in energy but also happens at a magnetic field strength that depends on the chain length. It thus becomes experimentally very hard to distinguish between the MBSs and a topological phase versus the trivial phase with its ABSs.

    \begin{figure}[t]
    \begin{center}
        \includegraphics[width=1\columnwidth]{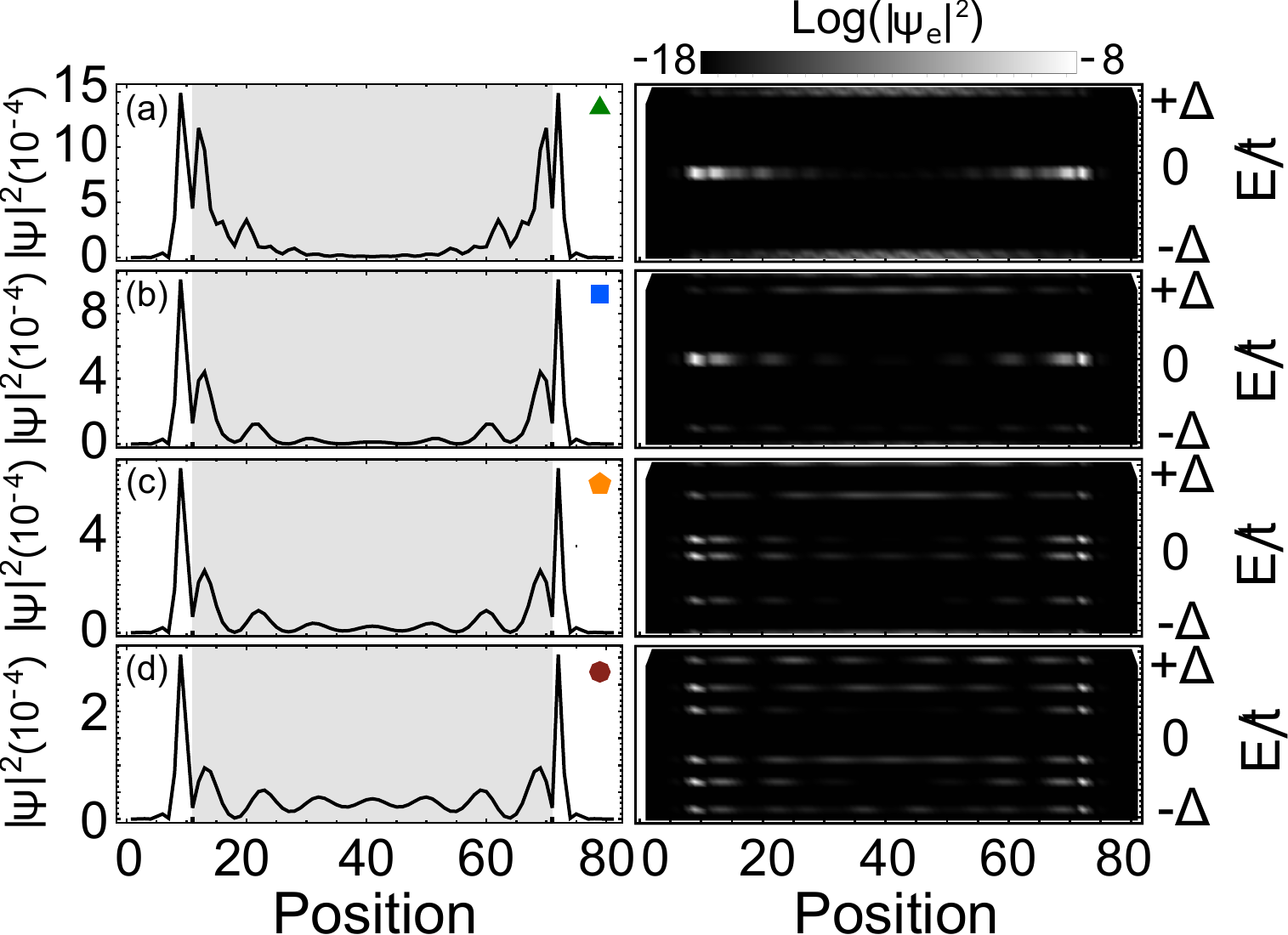}
        \caption{Left: Local-density profile $|\Psi(E_b,i)|^2$ (electron and hole components) of the state with energy $\pm E_b$ at site position $i$ for the lowest-lying states marked in Fig.~\ref{fig:oscillations_mu04_thetapi2}(c). Shaded gray region indicates the sites along the 61-site chain. Right: Electron component of the local density of states $|\Psi^{e}(E,i)|^2$ of subgap states ($-\Delta \leq E \leq +\Delta$). Each panel corresponds to a different value of $V_z$, as indicated by the corresponding symbol in Fig.~\ref{fig:oscillations_mu04_thetapi2}(c): (a) $V_z\!=\!10t$, (b) $V_z\!=\!15t$, (c) $V_z\!=\!17t$, and (d) $V_z\!=\!20t$.  The results show a smooth transition between MBSs in (a) and ABSs in (d).
        }
        \label{fig:stateprobability}
    \end{center}
\end{figure}

A further check to differentiate MBS and ABS is the local density of the lowest-lying states, at energy $\pm E_b$, in the Bogoliubov-de Gennes spectrum (both electron and hole contributions) at each site $i$ along the chain, i.e., ~$|\Psi(E_b, i)|^2$. Another important and closely related quantity, which can be accessible using local probe experiments such as STM setups, is the electronic local density of states for the (bound) states within the superconducting gap, which is calculated as $|\Psi^{e}(E,i)|^2$ as it takes into account only the electron contribution of the Bogoliubov-de Gennes spectrum. Figure \ref{fig:stateprobability} shows both quantities for a 61-site chain and for the $V_z$ values marked in Fig.~\ref{fig:oscillations_mu04_thetapi2}(c). These plots confirm the smooth evolution from MBSs to ABSs as $V_z$ approaches $V^{L=61}_{zc}$.
Deep in the topological regime ($V_z \ll V^{L=61}_{zc}$), the density profile along the chain is consistent with that expected for MBSs: The density is strongly localized at the ends of the chain and $|\Psi^{e}(E,i)|^2$ shows a large contribution at $E\!=\!0$, as shown in  Fig.~\ref{fig:stateprobability}(a). As for $V_z \gg V^{L=61}_{zc}$ in Fig.~\ref{fig:stateprobability}(d), the low-lying states instead show a clear ABS character, with $|\Psi(E_b,i)|^2$ being more delocalized along the entire chain and having large $|\Psi^{e}(E,i)|^2$ contributions at $E\!\neq\!0$, as well as a slight asymmetry in energies  $|\Psi^{e}(E,i)|^2 \! \neq \! |\Psi^{e}(-E,i)|^2$.Figures \ref{fig:stateprobability}(b) and \ref{fig:stateprobability}(c) illustrate the smooth MBS-ABS transition for $V_z \sim V^{L=61}_{zc}$.

    \begin{figure}[t]
        \begin{center}
            \includegraphics[width=1\columnwidth]{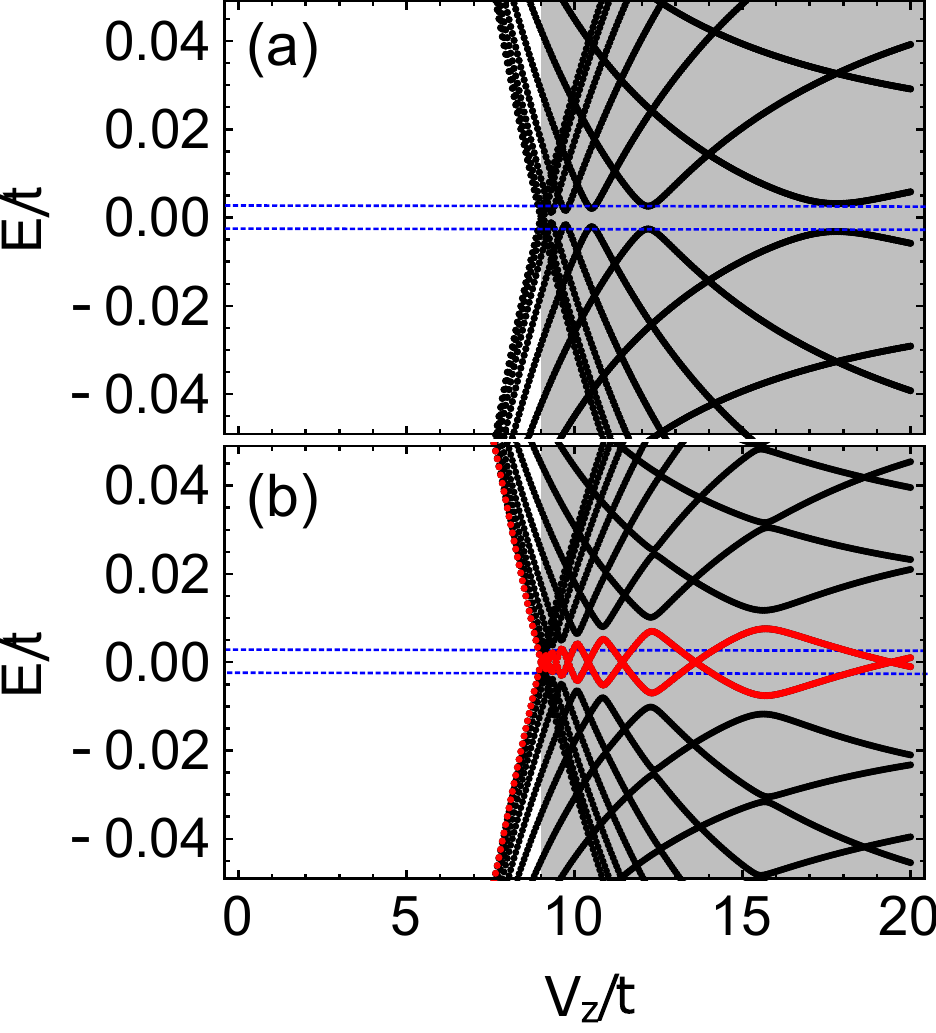}
            \caption{(a) Spiral $\theta=\pi/4$ chain with $\mu=0.2t$ with PBCs and (b) a fully embedded 25-site chain. States closest to zero energy are in red, horizontal dotted blue lines mark the value of the superconducting order parameter in the bulk $\pm \Delta$, and shaded regions mark the $V_z$ range of the topological phase ($\mathcal{M}=-1$) from Fig.~\ref{fig:phasediagrams}(c).
}
            \label{fig:oscillations_mu02_thetapi4_vslr}
        \end{center}
    \end{figure}
    For the cases when there is no upper bound in $V_z$ for the topological region it is even harder to make a clear distinction between MBS and ABS states. One such case is shown in Fig.\ \ref{fig:oscillations_mu02_thetapi4_vslr} for $\theta=\pi/4$ and $\mu=0.2t$. The corresponding phase diagram in Fig.\ \ref{fig:phasediagrams}(c) shows a lower bound for the topological region at $V_z \! \approx \! 8t$m but we find no upper bound up to $V_z\!=\!40t$. In this situation, the lowest-energy state's amplitude oscillations seen in Fig.\ \ref{fig:oscillations_mu02_thetapi4_vslr}(b) increase in amplitude, as predicted by the ansatz in Eq.\ (\ref{eq:ansatz}). There is, however, a clear \textit{increase} in the oscillation period, which is at odds with the predictions of Eq.\ (\ref{eq:ansatz}). Intriguingly, the overall amplitude can be of the order of the superconducting gap ($\Delta\approx0.004t$).
        Moreover, just inside the topological phase we find MBSs localized at the wire end points as anticipated, but even for moderately small $V_z$, we find the lowest-energy state instead being localized across the full chain, behaving as an ABSs instead. Thus, the same phenomenon of a moving phase boundary with wire length as in Fig.~\ref{fig:oscillations_mu04_thetapi2} is likely present here as well, although exact boundaries are not possible to establish due to the unbounded topological phase.
    
    Clearly, the results for spiral chains in Figs.~\ref{fig:oscillations_mu04_thetapi2_vslr} and \ref{fig:oscillations_mu02_thetapi4_vslr} show that one needs to exercise extreme caution in using an ansatz such as Eq.\ (\ref{eq:ansatz}) as a test for the presence or absence of MBSs in the spectrum as well as their properties. Depending on the microscopic details, the MBSs can show both increasing and decreasing oscillation amplitude and periods with increasing $V_z$ as well as a very smooth transition to ABSs within any finite-chain setup even if the infinite chain is in the topological phase. 
    
\begin{figure}[t]
        \begin{center}
            \includegraphics[width=1\columnwidth]{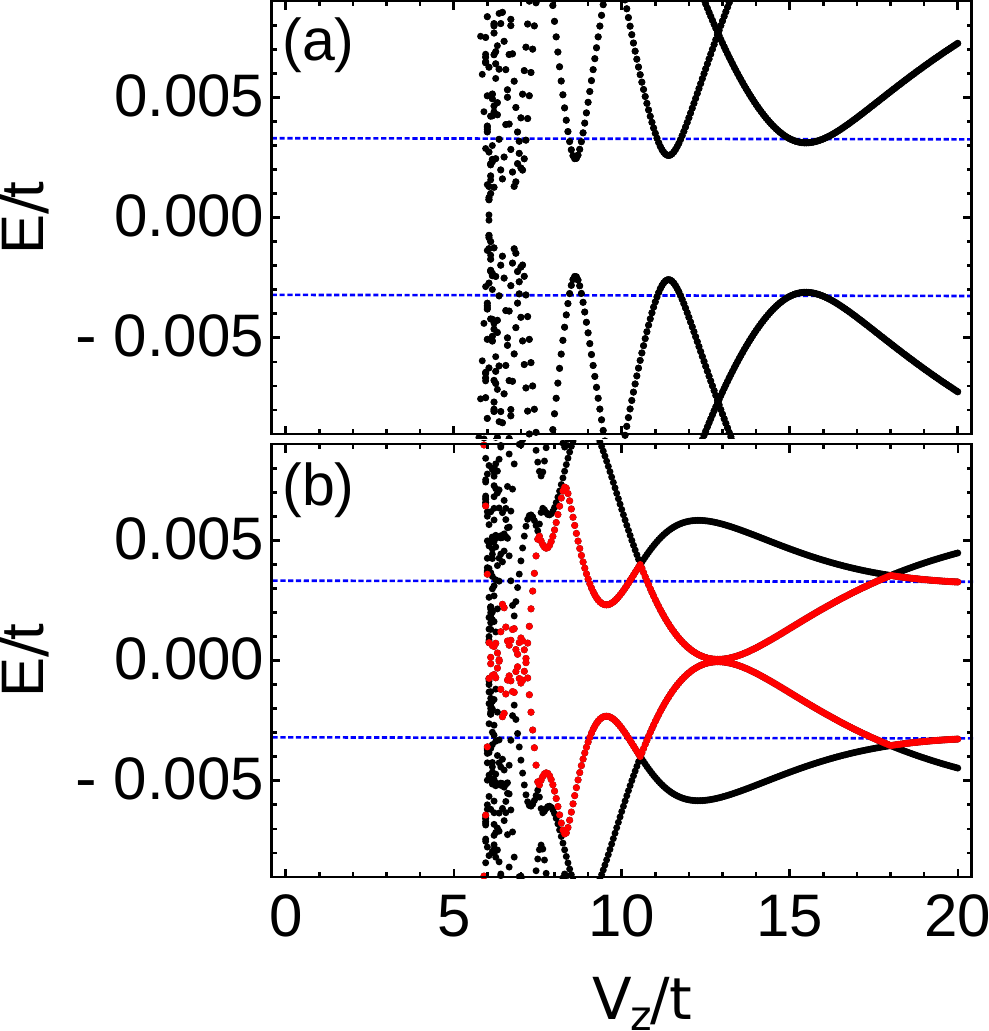}
            \caption{(a) Spiral $\theta=\pi/2$ chain with $\mu=0.192t$ with PBCs and (b) a fully embedded 21-site chain. States closest to zero energy are in red, and horizontal dotted blue lines mark the value of the superconducting order parameter in the bulk $\pm \Delta$. Here $\mu$ is chosen to induce a topologically trivial region for all $V_z$.}
            \label{fig:oscillations_ABS}
        \end{center}
    \end{figure}

    Finally, we consider the crossing points of the boundary curves in the $\theta \neq 0$ phase diagrams in Figs. \ref{fig:phasediagrams}(b) and  \ref{fig:phasediagrams}(c). These points correspond to values of $\mu$ where no topological phases appear independently of the value $V_z$; any low-energy state states should thus all be  ABSs. One example is $\theta=\pi/2$ and $\mu=0.192 t$, as shown in Fig.\ \ref{fig:oscillations_ABS}. The subgap states appearing in the finite-chain calculations [red points in Fig.\ \ref{fig:oscillations_ABS}(b)] are all ABSs and the gap oscillations are clearly distinct from those cases where MBSs appear. We have confirmed that these states are also not localized on the chain end points but spread over the whole chain. We notice also that the gap does not close completely (within numerical precision). 

    \section{Concluding remarks}
    \label{sec:Conclusions}
    To summarize, we studied the topological phases and MBSs in a chain of magnetic impurities on a honeycomb topological insulator with induced $s$-wave superconductivity. The shape of the doping vs magnetic field phase diagram changes significantly for different configurations of the magnetic ordering along the chain, showing features such as crossings of phase boundaries and unbounded topological regions depending on the spiral angle of the magnetic moments. 
    
    Importantly, we also showed that the effectiveness of using gap oscillations as a tool to distinguish topological Majorana bound states and non-topological Andreev bound states in this system is strongly impaired. In some cases, a behavior similar to that predicted in the literature for nanowires \cite{DasSarma:Phys.Rev.B:220506:2012}, with increasing amplitude and decreasing oscillation period with magnetic field, was obtained. In other cases, however, we obtained MBS-generated gap oscillations which behave completely differently, decreasing in amplitude and/or increasing in oscillation period, as well as similar oscillatory behavior in nontopological regions of the phase diagram, where only ABSs are present. We note that a comparison between the studied system and simple nanowires might not be done easily, as the models are quite different. However, since both belong to the same topological class we expect that a similarly varied behavior of the gap oscillations is also present in more elaborate models for nanowires. The difficulty in using gap oscillations to distinguish MBSs and ABSs is most pronounced in regions of the phase diagram without a clear upper bound in magnetic field for the topological phase. Here  the magnetic chain length also plays a fundamental role in the formation of MBSs or ABSs, even if the infinite chain is in the topologically nontrivial phase.
    
    In conclusion, our results show that topological phases and their associated MBSs can show wildly different behaviors even in very simple models of real materials. In particular, our results strongly caution against interpreting experimental results of oscillating low-energy states as indicative of non-trivial topology based on specific oscillations properties. Our results could potentially be verified using transport and local density of states measurements in a honeycomb QSHI material \cite{Haruyama:arXiv:2018,Wu:Science:76:2018}, with magnetic impurities, deposited on top of a conventional superconductor.

    \begin{acknowledgments}
        
        The authors are grateful to Uppsala University and the Institute of Physics of the University of S\~ao Paulo for the exchange program that allowed this project to be develop. R.L.R.C.T. and L.G.G.V.D.S. acknowledge support from FAPESP Grant No. 2016/18495-4, CNPq (graduate scholarship program, and Research Grants No.~308351/2017-7 and No.~449148/2014-9). D.K. and A.M.B-S. acknowledge support from the Swedish Research Council (Vetenskapsr\aa det, Grant No.~621-2014-3721), the Carl Trygger Foundation, the G\"{o}ran Gustafsson Foundation, the Swedish Foundation for Strategic Research (SSF), and the Knut and Alice Wallenberg Foundation through the Wallenberg Academy Fellows program.
        
    \end{acknowledgments}

\appendix

\section{Influence of next-nearest-neighbor hopping and spin-orbit asymmetry in the phase diagrams}
\label{app:SecondNeighbor}

     \begin{figure}[t]
        \begin{center}
            \includegraphics[width=1\columnwidth]{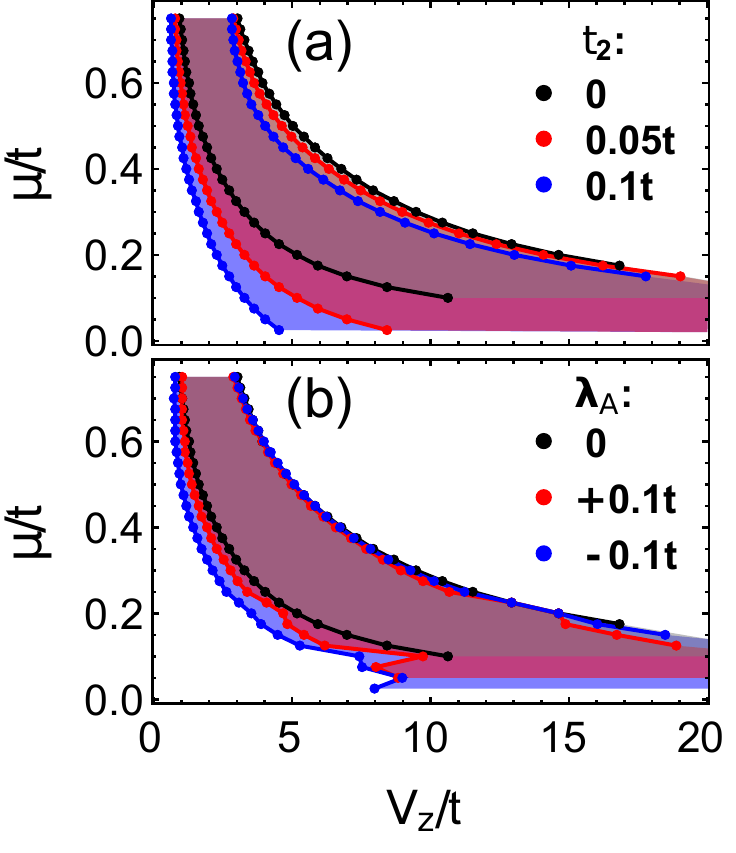}
            \caption{Phase diagrams for a ferromagnetic chain ($\theta=0$) with (a) next-nearest-neighbor hopping $t_2$ and (b) sublattice asymmetric spin-orbit couplings $\lambda_{A}/\lambda_{SO} =\pm 0.2$ or 20\% asymmetry. The data for $t_2\!=\!0$ and $\lambda_A\!=\!0$ are the same as presented in Fig.\ \ref{fig:phasediagrams}.}
            \label{fig:phasediagram_t2}
        \end{center}
    \end{figure}

In this appendix we consider the influence of two corrections to the Kane-Mele model Hamiltonian given by Eq.\ (\ref{eq.KM}), which might be present in real materials forming bipartite hexagonal lattices, namely, (i) adding a next-nearest-neighbor (NNN) hopping term $t_2$ and (ii) considering a sublattice asymmetry in the spin-orbit coupling.

Let us begin by considering a finite NNN term. As is wellknown, NNN hopping induces particle-hole symmetry (PHS) breaking in the band structure of bipartite honeycomb lattices \cite{castro-neto_rmp_2009}. 
Since MBSs are zero-energy modes protected by the symmetries guaranteeing the nontrivial topology, it is important to consider the stability of the phase diagram when breaking various symmetries. We note, however, that due to the finite chemical potential PHS is already broken in Eq.~\eqref{eq.KM}. Furthermore, including finite NNN, we find that it actually increases the topological regions in the phase diagrams. As shown in Fig.\ \ref{fig:phasediagram_t2}(a), the topological regions in the phase diagram for a FM chain are, in fact, \textit{enlarged} as the NNN hopping increases. Most interestingly, the lower phase boundary extends all the way down to $\mu\! \approx \!0$ for $t_2\!\neq\!0$. 
This latter effect we can attribute to the importance of breaking the PHS: At finite $\mu$ PHS is already broken, but at $\mu =0$ a nontrivial topological phase appears only for finite NNN hopping. As NNN hopping is certainly relevant to realistic experimental realizations, the results presented in Fig.\ \ref{fig:phasediagram_t2} indicate that the topological phases will be even more robust in realistic samples and more independent of the doping level.

The second effect is a sublattice asymmetry in the spin-orbit coupling $\lambda_{SO}$ in Eq.\ (\ref{eq.KM}). To investigate this, we implement a spin-orbit coupling of the form $\lambda^{A(B)}_{SO} = \lambda_{SO} \pm \lambda_A$, such that a nonzero $\lambda_A$ breaks the sublattice symmetry of the spin-orbit coupling. Figure \ref{fig:phasediagram_t2}(b) shows that even a large asymmetry of 20\% in the A,B spin-orbit couplings does not bring any significant changes to the phase diagram, apart from the topological region again stretching down to $\mu =0$ due to an effective PHS breaking.
  \nocite{Gonzalez:Phys.Rev.B:115327:2012}\nocite{He:Phys.Rev.B:075126:2013}\nocite{Inglot:JournalofAppliedPhysics:123709:2011}\nocite{Nadj-Perge:Phys.Rev.B:020407:2013}\nocite{Reis:Phys.Rev.B:085124:2014}
    
%

\end{document}